\documentclass[proceedings,article,submit,moreauthors]{mdpi} 
\firstpage{1} 
\pubvolume{xx}
\issuenum{1}
\articlenumber{5}
\pubyear{2019}
\copyrightyear{2019}
\history{Published: 16 September 2019}
\Title{Discovery of Jet Induced Soft Lags of XTE J1550-564 during its 1998 Outburst}

\Author{Arka Chatterjee $^{1,*,\ddagger}$\orcidA{}, Broja G. Dutta $^{2}$\orcidB{}, 
Dusmanta Patra$^{3}$\orcidC{}, Sandip K. Chakrabarti$^3$\orcidD{} and Prantik Nandi$^1$\orcidE{}}

\AuthorNames{Arka Chatterjee, Broja G. Dutta, Dusmanta Patra, Sandip K. Chakrabarti and Prantik Nandi}

\address{%
$^{1}$ \quad S. N. Bose National Centre for Basic Sciences, Salt Lake, Kolkata, 700106, India; arkachatterjee@bose.res.in (A.C); prantiknandi@bose.res.in (P.N)\\
$^{2}$ \quad Rishi Bankim Chandra College, Naihati, West Bengal, 743165, India.; brojadutta@gmail.com\\
$^{3}$ \quad Indian Centre for Space Physics, 43 Chalantika, Garia St. Rd., Kolkata, 700084, India; dusmanta@csp.res.in (D.P); sandip@csp.res.in (S.K.C)}

\corres{Correspondence: arkachatterjee@bose.res.in}

\firstnote{Presented at the meeting {\it Recent Progress in Relativistic Astrophysics}, 6-8 May 2019 (Shanghai, China).}
\secondnote{Speaker}

\abstract{X-ray time lags are complicated in nature. The exact reasons for complex lag spectra are yet to 
be known. However, the hard lags, in general are believed to be originated due to inverse Comptonization 
process. But, the origin of soft lags remained mischievous. Recent studies on ``Disk-Jet Connections" revealed
that the jets are also contributing in the X-ray spectral and timing properties in a magnitude which was more
than what was predicted earlier. In this article, we first show an exact anti-correlation between X-ray time lag 
and radio flux for XTE J1550-546 during its 1998 outburst. We propose that the soft lags might be generated 
due to the change in the accretion disk structure along the line of sight during higher jet activity.}


\begin{document}
\section{Introduction}
Accretion onto black holes are one of the most energy efficient process that occurs in our universe. During accretion, 
matter heats up via losing potential energy and radiates. The radiation can be detected throughout the entire
electromagnetic spectrum. The nature of such radiations are found to be varying over timescales ranging from sub-second
to few days. In X-ray regime, Quasi Periodic Oscillations or QPOs are found for most of the Galactic Black Holes (hereafter GBHs) by taking FFT of the 
observed light-curve and can be of different types (see \cite{Mo15}). The centroid frequency ($\nu_c$) of QPOs 
which varies between $0.01-20$ Hz are considered as Low Frequency QPOs. The origin of such QPOs are described by 
shock oscillation model (see \cite{cm00}) and also via Lense-Thirring precision model 
(\cite{FA05}). Phase/Time lags are computed by taking cross spectrum of two different energy bands of the
observed X-ray light curve (\cite{Mi88}). Hard lag of positive lag is found when harder photons delay over 
the soft/reference band. Soft/negative lag is produced when the hard photons reach the observer prior to the soft photons. Hard 
lags are most commonly interpreted by the inverse Comptonization \cite{P80} while soft lags were modeled by propagatory 
perturbation model (see \cite{BL99} and\cite{Li00}). Also, it was suggested that the hard X-ray 
which are reprocessed by the Keplerian disk could explain the soft lags found in case of GBHs and AGNs \cite{Pou99}.
Recently, the dependency of lag signs over the inclination angle of the disk was brought 
into light (see \cite{Du16} and \cite{van17}) where the high inclination GBHs are documented as 
more prone to exhibit soft lags. However, the connection between X-ray timing properties and radio 
jets became tighter when origin of type-B QPOs due to the oscillation of the base of the jet was 
recommended \cite{van17}.

Jets are one of the common mechanism via which a part of the accreting matter is ejected. In recent years, 
the studies of apparent superluminal jet launching mechanism (see \cite{Mi94} for details) from 
the disk has evolved from the point of observations. Earlier, the disk and jets are believed 
to be of different origin. X-ray flux ($F^X$) and radio flux ($F^R$) correlation for 
the entire mass range of black holes (\cite{Me03}; \cite{Co03}, \cite{Ca07}; \cite{SF11} 
and \cite{Co11}) suggested a strong connection between accretion disk and radio jets. 
Lag between X-ray and optical band of GX-339-4 \cite{Ga08} implicates the 
lag may have originated due to the modulation of magnetic field near the jet base. These studies raised questions
whether the time lag which is calculated by integrating over the $\nu_c \pm FWHM$ has any contribution from the jet. 
In presence of Comptonization, reflection and gravitational bending, 
complex lag properties of GBHs were examined \cite{Ch17b} where it was discussed 
that the outflows/jets should be a major component 
which could enhance the formation of soft lags. GRS 1915+105 had followed a pattern where 
it changed the lag sign during higher radio activity \cite{Mu01}. 

Keeping this disk-jet connections in mind, we studied XTE J1550-564 which was observed 
in multi-wavelengths. It was discovered by the All-Sky Monitor (ASM) on board Rossi X-Ray Timing 
Explorer (RXTE) in 1998 \citep{Sm98}. RXTE spectral observations of XTE J1550-564 
was reported \citep{So00} and an estimated mass of the black hole was found to be 
$M_{BH} \sim 10 M_{\odot}$. The source was observed in multi-wavelengths 
during its 1998 outburst \citep{Wu02}. The outburst begins with a hard X-ray spike, 
but the soft X-ray originated from the standard disk dominated in the later 
stages. \citep{Cu99} reported the discovery of strong QPOs from the X-ray 
light curves of XTE J1550-564. A high frequency QPOs near $160-215$ Hz  
was reported \citep{Re99}. In the rising state, significant hard lags were seen \citep{Cu00}. 

In this paper, we investigate radio flux and X-ray time lag of GBH XTE J1550-564. We provide 
data analysis process and results in the following sections. In \S 4, we discuss the origin of 
such correlation under Two Component Advective Flow (see \cite{CT95}) 
paradigm and we draw our conclusion in \S 5.


\begin{figure}[h!]
\vskip 0.5cm
{\includegraphics[height=0.5\columnwidth,width=1.1\columnwidth]{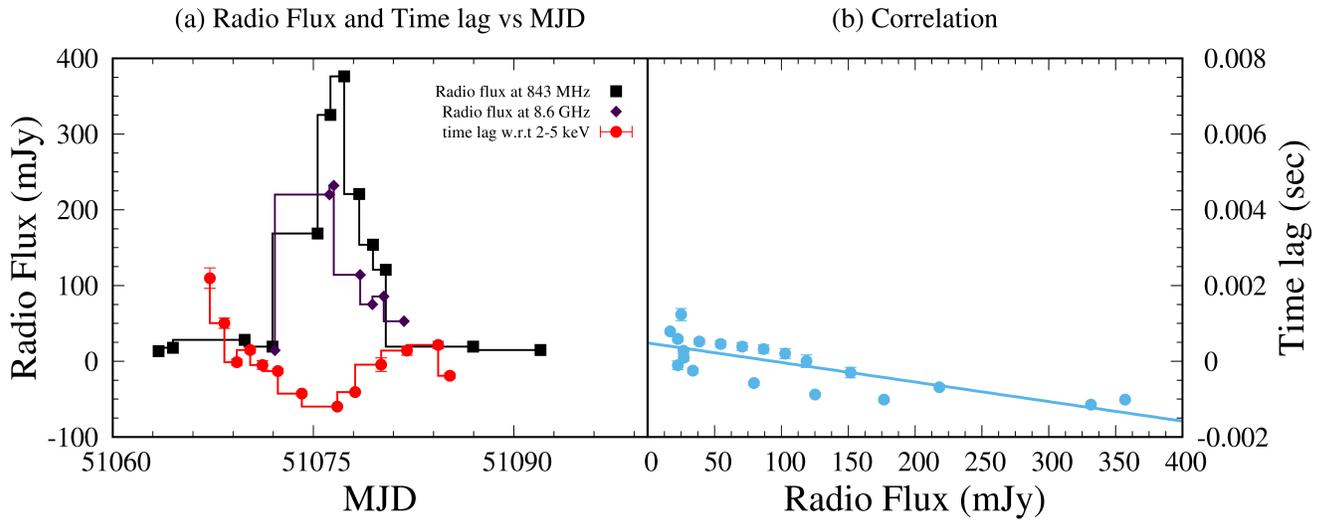}}\hskip 0.2cm
\caption{(a) Radio flux and X-ray lags are plotted with MJD for XTE J155-564. 
(b) Correlation between radio flux and time lag is plotted for XTE J1550-564. We opted for
near simultaneous approach for the radio and time lag data. The fitted slope ($a=-6.01\times10^{-6}$) is found to be 
negative.}
\label{}
\end{figure}
\section{Data analysis}
RXTE PCA archival data is used to generate lag spectra. Cross spectra are calculated. Phase
lag between two band signals at a Fourier frequency $\nu_j$ is given by $\phi_j = arg[CF(j)]$
and the corresponding time lag is $\phi_j/{2 \pi \nu_j}$ (see \cite{Ut14}). Lags are calculated
at the QPO centroid frequency ($\nu_c$) integrating over the interval $\nu_c \pm FWHM$
for 5-13 keV energy band against 2-5 keV energy band.

For XTE 1550-564, we have used the published radio flux at 843 MHz from MOST by \cite{Wu02} and 
8.6 GHz from ATCA radio telescope by \cite{Ha01} respectively.

\section{Results}
During the 1998 outburst, the source was observed in radio, optical 
and X-rays. The radio and optical counterparts of the source was repoted 
(see for details \cite{Ca98} and \cite{Or98}). RXTE monitored
the X-ray activity which maximized around MJD 51076. Subsequently, a giant
radio jet was observed.

The rising state of the outburst continued until MJD 51076. The QPO frequency 
increased. But, the associated lag changed the sign after MJD 51070. From Fig. 1a, 
we can see the increase in X-ray soft lags with the increase in radio flux. The 
correlation curve in Fig. 1b suggest the jet activity induced soft lags for this 
outburst. The Pearson Correlation Coefficient 
is $-0.754$ for this outburst. Here, we have reported only up to MJD 51093 as the QPOs
after that became sporadic in nature and simultaneous radio observations were absent. 
Also, X-ray spectral studies of XTE J1550-564 was carried out where 
the absence of QPOs are reported after MJD 51150 and the X-ray spectra was found to 
be highly disk dominated in the $2-20$ keV range \cite{So00}. 

\section{Discussion}
\begin{figure}[h!]
\vskip 0.5cm
{\includegraphics[height=0.30\columnwidth,width=0.95\columnwidth]{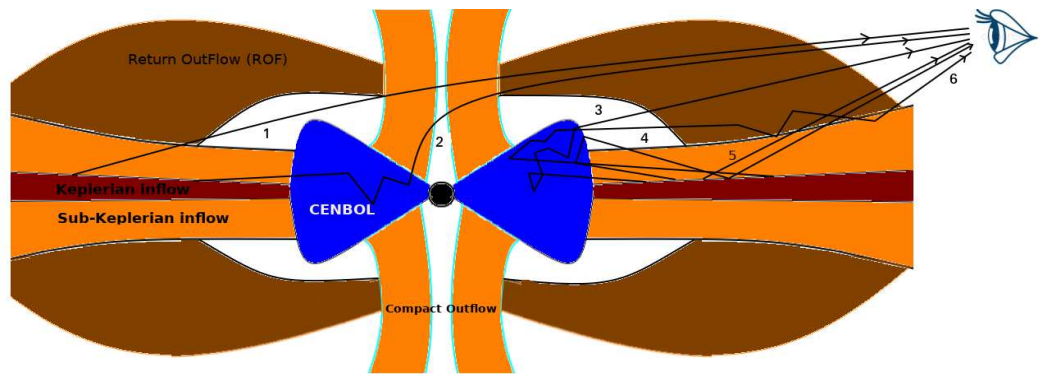}}\hskip 0.2cm
\caption{Cartoon diagram of Two Component Advective Flow (TCAF) in presence of Return OutFlow
(ROF) is shown. Pseudo colors represent approximated temperature profile within the disk. Six emergent photon types
are shown in the cartoon. Photon type (1) denotes the soft photons which have suffered gravitational bending but are 
not intercepted by the CENBOL. Photon (2) represents the hard photons which are scattered in the CENBOL
and undergone gravitational bending to reach the distant observer at high inclination angle. Photon
(3) scattered in the CENBOL and without suffering bending reach the observer. The fourth category of photons 
belong to the hard photons which are reflected by the Keplerian disk. Number five represents 
the soft photons which arrived the detector plane without scattering. Photon type (6) denotes 
the photons which scattered in the Compton cloud and again downscattered in the ROF region 
while reaching the observer.}
\label{}
\end{figure}

QPOs and associated phase or time lags are one of the major information carrier of accretion disk geometry 
as well as their evolution during outbursts. The QPO centroid frequency ($\nu_{c}$) directly correlates with 
size of the Compton cloud (see \cite{cm00} for further details). On the other hand, lags 
are associated with fluctuation of thermodynamical parameters like density, temperature, reflection coefficient 
of the Keplerian disk and interception fraction of the soft photons (see \cite{Ch17b} 
for further details). Recent studies of spectral (see \cite{Gh11}, \cite{Ch17a} and \cite{He15} for further details)
and temporal (see \cite{Mo15}) variabilities showed significant spectral hardening 
and QPO $\nu_{c}$ and/or {\it rms} variation due to inclination angle variation. Later, similar 
inclination dependency of lag signs were seen (see \cite{Du16} and \cite{van17}). It was shown 
that the soft lags are mostly found for the GBHs which are at high inclination angle.

XTE J1550-564 had undergone an outburst during 1998 where the X-ray flux has reached $6.8$ crab. 
During this time, the source had also shown a giant radio flare of $\sim 376$ mJy (see \cite{Wu02}). 
Associated lags also evolve with $\nu_c$ and radio flux. From Fig. 1a, we can see an anti-correlation 
between time lags and radio flux. The Fig. 1b suggests a strong correlation between radio 
flux and soft lags.

Under Two Component Advective Flow (TCAF) paradigm (see \cite{CT95} for further details), 
we investigate the physical origin of such correlation. The advective flow contains a sub-Keplerian component
and a Keplerian disk. Reaching the centrifugal barrier, the flow undergoes a shock which slows down the 
inflowing matter. This causes a sudden rise of temperature which puffs up the matter creating a hotter region 
known as CENBOL which is responsible for the inverse Comptonization of the soft photons generated by the 
Keplerian disk. Using hydrodynamic simulations (see \cite{GC12} and \cite{Ch18}), one finds self-consistent 
outflows are being produced from the post-shock region. In our model, the QPOs are generated due to the shock oscillation 
at the CENBOL boundary. During accretion, a fraction of inflowing matter forms outflow. As the spectral state 
changes from hard to intermediate states, the amount of outflow increases and so does the $\nu_c$. A part of 
the outflowing matter which fails to achieve the escape velocity, returns to inflow causing the formation of 
a region cooler, yet denser than the rest of the inflow. The scattered hard photons passes through this region 
to reach the observer at high inclination angle. While passing through this Return OutFlow (ROF) region, a fraction 
of the hard photons downscatter. These downscattered photons (type 4 in Fig. 2) lags behind rest of the hard 
radiation (type 3 \& 3 in Fig. 2) emitted from Compton cloud. We argue that the origin of soft lag could be 
explained by the downscattering of the hard radiation in the ROF region. Similar correlations for other sources at high 
inclination angle were found and reported \cite{Pa19}.

\section{Conclusions}
{We report a correlation between radio flux and X-ray time lag for XTE J1550-564 during its 1998 
outburst from which we can conclude that the soft lags are simultaneous with the radio flux.
Our finding provides a model independent insight to the disk-jet connection and implicates 
the severity of simultaneous broadband studies of astrophysical black holes to be able to 
acquire deeper understanding.}

\vspace{6pt} 
\authorcontributions{A.C simulated the lag variation in presence 
of outflows and wrote the manuscript. B.D.G and D.P analyzed the data. S.K.C supervised the 
project and modified the manuscript. P.N provided the correlation data.} 

\funding{The work of A.C is funded by Advanced Post Doctoral Manpower Programme (APMP), 
S. N. Bose National Centre for Basic Sciences, Kolkata, India. The work of P.N is supported 
by CSIR, India.}

\acknowledgments{Authors thank Tomaso Belloni for providing the timing analysis software 
GHATS. AC acknowledges Kinwah Wu for valuable discussions. BGD acknowledges IUCAA 
associateship.}

\conflictsofinterest{The authors declare no conflict of interest.} 
\reftitle{References}

\end{document}